\documentclass{article}
\usepackage[super]{cite}
\usepackage{graphicx}
\usepackage{amsmath}
\usepackage{amsfonts}
\usepackage{geometry}

\newcommand{\beq}{\begin{equation}}
\newcommand{\eeq}{\end{equation}}

\newcommand{\be}{\begin{eqnarray}}
\newcommand{\ee}{\end{eqnarray}}

\renewcommand{\d}{\mbox{${\rm d}$}} %d differenziale non corsivo in math mode

\newcommand{\gn}{G_{\rm N}}

\title{\bf Covariant singularities: a brief review}

\author{Roberto~Casadio$^{ab}$\thanks{E-mail: casadio@bo.infn.it},
$\ $
Alexander~Kamenshchik$^{abc}$\thanks{E-mail: kamenshchik@bo.infn.it},
$\ $
and
Iber\^e Kuntz$^{de}$\thanks{E-mail: kuntz@fisica.ufpr.br}
\\
\\
\\
$^a${\em Dipartimento di Fisica e Astronomia, Universit\`a di Bologna}
\\
{\em via Irnerio~46, 40126 Bologna, Italy}
\\
\\
$^b${\em I.N.F.N., Sezione di Bologna, I.S.~FLAG}
\\
{\em viale B.~Pichat~6/2, 40127 Bologna, Italy}
\\
\\
$^c${\em L.D.~Landau Institute for Theoretical Physics}
\\
{\em of the Russian Academy of Sciences} 
\\
{\em 119334 Moscow, Russia}
\\
\\
$^d${\em Departamento de F\'isica, ICE}
\\
{\em Universidade Federal de Juiz de Fora}
\\
{\em Campus Universitário, Rua José Lourenço Kelmer, s/n}
\\
{\em Juiz de Fora -- MG}
\\
{\em 36036-900, Brazil}
\\
\\
$^e${\em Departamento de Física, Universidade Federal do Paraná}
\\
{\em PO Box 19044}
\\
{\em Curitiba -- PR}
\\
{\em 81531-980, Brazil}
}

\date{}

\begin{document}

%\markboth{R.~Casadio, I.~Kuntz and A.~Kamenshchik}
%{Covariant singularities: a brief review}

%%%%%%%%%%%%%%%%%%%%% Publisher's Area please ignore %%%%%%%%%%%%%%
%\catchline{}{}{}{}{}
%%%%%%%%%%%%%%%%%%%%%%%%%%%%%%%%%%%%%%%%%%%%%%%%%%%%%%%%%%%%%%%%%%%

\maketitle

%\pub{Received (Day Month Year)}{Revised (Day Month Year)}

\begin{abstract}
The Hawking-Penrose theorem is not covariant under field redefinitions. Should the invariance under such transformations be a true principle in Nature, spacetime singularities become dubious objects. We here review the concept of covariant singularities, that is, singularities that are invariant under both spacetime diffeomorphisms and field redefinitions.

%\keywords{singularity; field redefinition; effective action.}
\end{abstract}

%\ccode{PACS Nos.: include PACS Nos.}

\section{Introduction}
\setcounter{equation}{0}
\label{Sintro}
A major issue in gravitational physics is the prediction of singularities. The most common approaches to such a problem comprise modifications in the gravitational sector or the particle content of the theory, so as to falsify one of the premises of the Hawking-Penrose theorem \cite{Hawking:1973uf}. Because singularities take place at Planckian energies, it is a widespread opinion that some quantum theory of gravity ought to exist to induce the aforementioned modifications. Even though such a theory remains unknown, investigations of structural properties of quantum field theory might shed some light on the singularity problem. The invariance under field redefinitions is one such property \cite{Casadio:2020zmn,Casadio:2021rwj}.

Field redefinitions play prominent roles in physics. They are primarily used at the linear and perturbative level in high-energy physics \cite{Arzt:1993gz,Georgi:1991ch,Criado:2018sdb,Passarino:2016saj}, but their non-linear generalizations are being met with increasing interest, particularly in the study of gravity~\cite{Starobinsky:1979ty,Starobinsky:1980te,Bezrukov:2007ep,Barvinsky:2008ia,Calmet:2017voc,Solodukhin:2015ypa,Kazakov:1987ej,Mohammedi:2013oca,
Slovick:2013rya,Apfeldorf:1994av}. 
Fields are dummy variables in the path integral, hence it is generally expected that quantities computed from functional integrals, hence physics, should remain invariant under field reparametrizations. The calculation of off-shell observables (e.g. in-in correlation functions) is, however, usually performed with functional generators, such as the partition function and the effective action. These generators are defined by the coupling with an external source or background field and this coupling is not invariant under field redefinitions.

In the background field method, this non-invariance of the standard effective action is intimately tied to its gauge-fixing dependence \cite{Vilkovisky:1984st}. Needless to say, this leads to several problems, particularly with the interpretation of the renormalization group equations and with off-shell correlation functions of the kind that is measured in cosmology, for instance. It is thus reasonable to conjecture the principle of invariance under field redefinitions and perform modifications on the functional generators in order to obey such a principle. From this viewpoint, fields are coordinates in the infinite-dimensional configuration space, field redefinitions are changes of coordinates in this space, and path integrals take on a more geometrical taste, generalizing the usual theory of integration on manifolds. This has led to the Vilkovisky-DeWitt effective action \cite{Vilkovisky:1984st,DeWitt:1988dq,Ellicott:1987ir,Parker:2009uva}, where the coupling to the background field is made invariant by the introduction of a Levi-Civita connection in configuration space. The theory of path integration, however, still lacks a complete and rigorous mathematical understanding, thus depending on a particular procedure for its calculation, e.g.~discretization. Otherwise it only makes sense at a formal level.

In this review, we shall take the position that physics at the fundamental level
should not depend on the way fields are parameterized.
There are at least three important reasons for this:
\begin{itemize}
	\item[(i)] as explained above, dependence on the field parametrization also leads to gauge-fixing dependence, thus the latter is automatically solved with a theory invariant under field redefinitions;
	\item[(ii)] the classical action does not depend on the field parameterization, thus it is reasonable to keep this property in the quantum theory as well;
	\item[(iii)] no particular parameterization is, in principle, favored. Giving special meaning to one such choice thus seems rather artificial. Indeed, experimental results do not imply any specific field parameterization. 
\end{itemize} 
\par
The paper is organized as follows: in Section~\ref{Sgeo}, we introduce the concept of covariant singularities. There are two subclasses of these singularities, one that results from singularities in the configuration-space geometry (Section~\ref{geosing}) and another that takes place directly in the observables (Section~\ref{Ssingularity}). The latter is of topological nature and can be characterized by a non-vanishing winding number. In Section~\ref{SecEx2}, we apply the formalism to a simple example in cosmology, where a spacetime singularity exists but turns out not to correspond to singular observables. We then draw our conclusions in Section~\ref{Sconc}.
\par
The presence of many types of indices and spacetime dependence makes it useful to adopt DeWitt's condensed notation
(for more details, see Ref.~\cite{Parker:2009uva}).
Mid-alphabet Greek letters (e.g.~$\mu,\nu,\rho, \ldots$)
shall thus denote spacetime indices, as usual;
lowercase mid-alphabet Latin indices (e.g.~$i, j, k, \ldots$) collectively represent
both discrete indices (denoted by the corresponding capital Latin letters $I, J, K, \ldots$),
and the continuum spacetime coordinates $x\equiv x^\mu$.
This correspondence can be formally written as $i = (I, x)$, thus $\phi^i = \phi^I(x)$
are the coordinates of a field configuration.
Repeated mid-alphabet lowercase indices result in summations over
all the discrete indices and integration over the spacetime $\Omega$
of dimension ${\rm dim}(\Omega)=n$.
Lowercase Latin indices of the beginning of the alphabet (e.g.~$a, b, c, \ldots$)
shall correspond to gauge indices, whereas indices from the beginning of the Greek alphabet (e.g.~$\alpha, \beta, \gamma, \ldots$)
will be reserved to spinor indices.

\section{Covariant Singularities}
\label{Sgeo}
\setcounter{equation}{0}
When the principle of field-redefinition invariance is postulated, many known results in physics should be reexamined. There is no guarantee that established results remain invariant under field redefinitions. One such example is the Hawking-Penrose theorem, whose formulation does not transform covariantly under field redefinitions~\cite{Casadio:2020zmn}. Singularities present for certain configuration-space coordinates might therefore not correspond to singular points in the field space parameterised in some other coordinates, in very much the same way that the event horizon is singular in the standard Schwarzschild spacetime coordinates but not in other coordinates. We thus define covariant singularities as the ones that are invariant under both changes of spacetime coordinates and field redefinitions.

\subsection{Configuration-space Singularities}
\label{geosing}
One type of covariant singularities is given by the ones appearing in curvature invariants of the configuration space~\cite{DeWitt:1967yk,Isham:1975ur,Giulini:1993ui,Giulini:2009np,Giulini:1994dx,Giulini:1993ct}.
If one adopts a Riemannian (or pseudo-Riemannian) structure for the configuration space, one needs to specify a configuration-space metric.
Such a metric, hereby denoted $G_{ij}$, must be seen as part of the definition
of the theory, along with the classical action.
The line element is then defined as usual as
\beq
\d\mathfrak{s}^2
=
G_{ij}\,\d\phi^i\, \d\phi^j
=
\int_\Omega\d^nx
\int_\Omega\d^nx'
\,G_{IJ}(x,x')\,\d\phi^I(x)\,\d\phi^J(x')
\ .
\eeq
One must furthermore require that $G_{ij}$ be invariant under the same gauge symmetries of the classical action.
This is important in order to enforce these symmetries at the quantum level via the path integral measure, which takes a factor $\sqrt{\det G_{ij}}$ to cancel
out the Jacobian determinant from the field redefinition, thus preventing gauge anomalies.
Apart from symmetry, we require ultralocality 
\begin{equation}
	G_{ij} = G_{IJ} \,\delta(x,x')
\end{equation}
where $G_{IJ}$ depends only on the fields $\phi^I$ but not on their derivatives. Ultralocality is particularly important to make contact with the theory of scattering amplitudes. There still remains to be determined the metric $G_{IJ}$ defined on the finite-dimensional subspace of the configuration space with fixed spacetime point $x^\mu =$ const.
We can adopt the spirit of effective field theory to organize all the infinite possible terms in the $G_{IJ}$ according to their mass dimensions. We shall then focus on the leading contribution, which contains only dimensionless parameters.

For metric theories of gravity, one identifies $\phi^I(x) = g^{\mu\nu}(x)$.
The assumption of simplicity, together with the symmetries of $G_{IJ}$,
then leads to the two-parameter family of field-space metrics~\cite{DeWitt:1967ub}
\beq
G_{ij}
=
\frac{1}{2}\,(-g)^\epsilon 
\left( g_{\mu\rho}\, g_{\sigma\nu}
+ g_{\mu\sigma}\, g_{\rho\nu}
+ c \, g_{\mu\nu}\, g_{\rho\sigma}
\right)
\delta(x,x') 
\ ,
\label{dwgen}
\eeq
with $g = \det g_{\mu\nu}$, which involves only the dimensionless parameters $c$ and $\epsilon$.
The coefficients of the first two terms in Eq.~\eqref{dwgen} are determined by
requiring that $G_{IJ}$ is a spacetime tensor that satisfies the invertibility condition $G_{IJ} \,G^{JK} = \delta^K_I$.

The connection in configuration space is then assumed to be of the Levi-Civita type:
\begin{equation}
\Gamma^i_{\ jk}
=
\frac12 \,G^{il}
\left(\partial_j G_{kl} + \partial_k G_{jl} - \partial_l G_{jk}
\right)
\end{equation}
and the functional Riemannian tensor is defined in the usual way
\begin{equation}
\mathcal R^i_{\ jkl}
=
\partial_k \Gamma^i_{\ lj} 
- \partial_l \Gamma^i_{\ kj} 
+ \Gamma^i_{\ km}\,\Gamma^m_{\ lj} 
+ \Gamma^i_{\ lm}\,\Gamma^m_{\ kj}
\ ,
\end{equation}
with $\mathcal R_{jl} = \mathcal R^i_{\ jil}$ and $\mathcal R = \mathcal R^i_{\ i}$ being
the functional Ricci tensor and functional Ricci scalar, respectively.
Note that the assumption of ultralocality implies that many contractions
will diverge as $\delta(x,x)$.
This only reflects the infinite dimension of the configuration space and can be easily amended by defining densities,
such as
\begin{equation}
	\frac{\mathcal R_{ijkl}\, \mathcal R^{ijkl}}{{\rm dim}(\mathcal M)}
	\propto
	\int\d^n x\, R_{IJKL}\, R^{IJKL}
	\ ,
\end{equation}
where $\mathcal M$ denotes the configuration space and ${\rm dim}(\mathcal M) = G_{ij} \,G^{ij}$ its dimension. Therefore, one good way to reveal the presence of singularities is through the calculation of the functional Kretschmann scalar
\beq
\mathcal K
=
\mathcal R_{IJKL}\,\mathcal R^{IJKL}
\ .
\label{functk}
\eeq
Since Eq.~\eqref{functk} is invariant under both spacetime diffeomorphisms and field redefinitions, a singular field configuration in $\mathcal K$ would signal the existence of a singularity.
In particular, the Kretschmann scalar for the functional metric \eqref{dwgen} reads~\cite{Casadio:2020zmn}
\beq
\mathcal K
=
\frac{n}{8}\, (-g)^{-2\epsilon}
\left(
\frac{n^3}{4} + \frac{3\,n^2}{4} - 1
\right)
\ .
\label{eq:general}
\eeq
Different choices for $\epsilon$ are in principle possible~\cite{Hamber:2009zz}, with $\epsilon = 0$ originally proposed by Misner~\cite{Misner:1957wq} and $\epsilon=1/2$ by
DeWitt~\cite{DeWitt:1962ud,DeWitt:1967yk,DeWitt:1965jb}.
Notice that apart from the case $\epsilon=0$, where $\mathcal K$ is independent of the spacetime metric, there is a covariant singularity at either $g = 0$ or $|g|\to\infty$.
In any case, it is not clear whether a singularity in $\mathcal K$ would be a real issue.~\footnote{We recall that the spacetime Kretschmann scalar
diverges for physically acceptable integrable singularities in which tidal forces remain finite.}
 Since covariant singularities belong to the boundary of the configuration space, and not to the configuration space itself, singular configurations cannot be solutions to the theory.
\par
It is also not clear how singularities in the configuration-space geometry affect the physical observables. At the quantum level, quantities of interest result from the interplay of the configuration-space geometry, the classical action, and the boundary conditions, thus assessing only the first of these is not sufficient.
\subsection{Functional Singularities}
\setcounter{equation}{0}
\label{Ssingularity}
A more transparent study of covariant singularities should be made in terms of physical observables. A natural candidate for this investigation is the effective action, where it is contained all observables of a quantum field theory.
As we have pointed out in the Introduction, the Vilkovisky-DeWitt effective action is a modified version of the standard effective action that incorporates the configuration-space geometry in its definition so as to preserve the invariance under field redefinitions. By replacing functional derivatives by their covariant counterparts and differences by distances along geodesics in configuration space, one is then able to couple the background field with the quantum field in an invariant manner:~\footnote{We set $\hbar = 1$ for simplicity.}~\cite{DeWitt:1988dq}
\beq
\exp\left\{i\,\Gamma[\varphi]\right\}
=
\int\d\mu[\phi]\,
\exp\left\{
i
\left(
S[\phi]
-\sigma^i(\varphi,\phi)\,(C^{-1})^j_{\ i}[\varphi]\,\nabla_{j}\Gamma[\varphi]
\right)
\right\}
\ ,
\label{eq:qadef}
\eeq
where
\beq
\sigma^i(\varphi,\phi)
=
\frac12
\left(\int_{\gamma(\varphi,\phi)}\d\mathfrak{s}\right)^2
\eeq
is the geodetic interval (analogous to Synge's world function~\cite{J.L.Synge:1960zz}),
calculated along the geodesic $\gamma$ with end-points
$\varphi$ and $\phi$, and $C^i_{\ j} = \langle\nabla_j\sigma^i(\varphi,\phi)\rangle_T$.
The angular brackets here denote the functional average, which, for any functional
$F[\varphi,\phi]$, is given by
\beq
\langle F[\varphi,\phi]\rangle_T
=
\exp\left\{-i \,\Gamma[\varphi]\right\}
\int\d\mu[\phi]\,
F[\varphi,\phi]\,
\exp\left\{
i
\left(S[\phi] + T^i[\varphi,\phi]\, \nabla_i\Gamma[\varphi]
\right)
\right\}
\ ,
\eeq
where $T^i[\varphi,\phi] = \sigma^i(\varphi,\, \phi)(C^{-1})^j_{\ i}[\varphi]$.
Note that $C^i_{\ j}$ is defined recursively since $C^i_{\ j}$ shows up in the functional average as well.
Solving for $C^i_{\ j}$ is clearly not easy and one usually resorts to an expansion in power series.
Being $\Gamma[\varphi]$ invariant under field redefinitions, we can define a covariant singularity as a solution 
$\varphi=\varphi_0$ to the effective equations of motion such that the Vilkovisky-DeWitt effective action $\Gamma[\varphi_0]$
evaluated at that point is not well-defined.
Contrary to the singularity in the configuration-space geometry defined before, the covariant singularity $\varphi_0$ does belong to the configuration space $\mathcal M$. Such a singularity thus
corresponds to an existing configuration with undefined observables. Some of these covariant singularities can, however, be removed by local alterations in the effective action without affecting its continuity in the far region. One example of such a procedure is the definition of $\Gamma[\varphi_0]$ as the limit of $\Gamma[\varphi]$ when $\varphi$ approaches $\varphi_0$. When such a procedure cannot be performed, the covariant singularity is not removable.
To such a non-removable covariant singularity we shall reserve the name of {\em functional singularity}.

Functional singularities cannot be removed without altering the global aspects of the theory. They indeed affect physical configurations arbitrarily far away in the configuration space. Their details, however, depend on the full knowledge of the effective action, thus are difficult to come by.
Fortunately, topological techniques come to our rescue in that they can provide tools to infer the presence of functional singularities. 

Functional singularities can be related to the topology of maps between the configuration space
and the real circle $\mathbb S^1$. Indeed, as suggested by the LHS of Eq.~\eqref{eq:qadef}, it is natural to define the functional order parameter~\footnote{This nomenclature is reminiscent of the study of topological defects in condensed matter.}
\beq
\psi[\varphi]
=
e^{i\,\Gamma[\varphi]}
\ ,
\label{orderp}
\eeq
to investigate functional singularities.
Points where the functional order parameter is singular correspond to configurations where the effective action is undefined, hence to functional singularities.
Functional singularities thus play a role analogous to topological defects in condensed matter~\cite{Mermin:1979zz}.
Assuming that $\Gamma$ is real,~\footnote{The functional order parameter space corresponding to complex effective actions is simply connected, thus its fundamental group is trivial and no functional singularity exists.}
the functional order parameter $\psi$ defines the map
\beq
\psi : \mathcal M \to \mathbb S^1,
\eeq
from the configuration space to the unit circle.
Should we encircle an exact solution $\varphi_0$
with a $d$-dimensional hypersurface $\gamma_d(\varphi_0) \subset \mathcal M$ with the topology of $\mathbb S^d$,
the functional order parameter restricted to $\gamma_d(\varphi_0)$ induces the map
\beq
\psi|_{\gamma_d} : \mathbb S^d \to \mathbb S^1
\eeq
between higher-dimensional spheres centered at $\varphi_0$ and the circle. The number of times $\mathcal W$ that $\psi|_{\gamma_d}$ wraps around $\mathbb S^1$ determines the nature of the functional singularity, with each value of $\mathcal W$ corresponding to topologically distinct cases. The functional singularity is absent (or removable) if, and only if, $\mathcal W=0$.
The topology of maps, hence the study of functional singularities, is characterise via the homotopy groups $\pi_d(\mathbb S^1)$, which encode the topologically different ways of wrapping $\psi|_{\gamma_d}$ around the unit circle. The interested reader can consult Refs.~\cite{Mermin:1979zz,Nakahara:2003nw} for more details on homotopy groups for physicists.
\par
Luckily, higher homotopy groups of the circle are all trivial, that is $\pi_d(\mathbb S^1) = \emptyset$ for $d>1$. The information on functional singularities is therefore fully contained in the fundamental group $\pi_1(\mathbb S^1) = \mathbb{Z}$,
which is isomorphic to the integers. These integers precisely represent the number of turns $\mathcal W$ defined above. Because configurations with $\mathcal W\neq 0$ are topologically distinct from $\mathcal W=0$, one cannot remove functional singularities by local and continuous alterations in the effective action, as we have already pointed out. In other words, one cannot continuously deform the loop $\gamma_1$, encircling a functional singularity, to a point. Notice that functional singularities are, in general, higher-dimensional subspaces of the configuration space and not points of zero dimensions.

The number of turns $\mathcal W$ is precisely accounted by the winding number:
\begin{align}
\mathcal{W}
&=
\frac{1}{2\,\pi\,i}
\oint_{\psi[\gamma_1]} 
\frac{\delta\psi}{\psi}
\nonumber
\\
&=
\frac{1}{2\,\pi}
\int_0^{2\,\pi}
\d\theta
\int_\Omega\d^n x\, 
\left.\frac{\partial \mathcal L(x)}{\partial \varphi^I(x)}\right|_{\varphi^I(x)=\gamma^I(x;\theta)}
\frac{\d\gamma^I(x;\theta)}{\d\theta}
\ ,
\label{wind2}
\end{align}
where $\psi[\gamma_1]$ denotes the image of $\gamma_1$ under the map
$\psi[\varphi]$.
The field configurations $\gamma^i=\gamma^I(x;\theta)$ are an explicit parameterisation
of $\gamma_1$ in terms of the angle $0\le\theta\le 2\,\pi$ such that
$\gamma^I(x;0)=\gamma^I(x;2\,\pi)$ and, of course,
\be
\Gamma[\varphi]
=
\int_\Omega
\d ^n x\,\mathcal{L}(\varphi^I,\partial_\mu\varphi^I,\ldots)
\ ,
\ee
with $\mathcal{L}$ the effective Lagrangian density.
We should stress that, since $\delta\Gamma=\delta\psi/\psi$ is an exact form, the winding number is independent
of the curve $\gamma_1$.
Finally, since $\mathcal W \neq 0$ is a necessary and sufficient condition for the presence of a functional singularity, Eq.~\eqref{wind2} provides a direct procedure to determine the extent to which a theory is well-defined.
\subsection{\textit{Example: scalar field in cosmology}}
\label{SecEx2}
In this subsection, we will illustrate the formalism introduced above for a theory of a massless scalar field $\phi$ minimally coupled to general relativity.~\footnote{We refer the reader to Ref.~\cite{Casadio:2021rwj} for more details on this calculation.}
For this purpose, we shall assume that the effective action is known and given by the simple expression:~\footnote{Because we shall only consider the homogeneous case, integration over space will produce an infinite volume. The notation $\tilde\Gamma$ is just a reminder that this action is IR divergent. We then define the IR-finite action $\Gamma$, without the tilde, to be the ratio between $\tilde\Gamma$ and the infinite volume.}
\beq
\tilde \Gamma 
=
\int_\Omega
\d^4x \, \sqrt{-g}
\left(
	\frac{R}{16\, \pi \,\gn}
	- \frac12\, \partial_\mu \phi \,\partial^\mu \phi
\right),
\label{eaS}
\eeq
where $\gn$ denotes Newton's constant and $R$ is the spacetime Ricci scalar. In real situations, the effective action is much more complicated (if calculable exactly at all) and it is usually non-local or non-analytic. One can also think of Eq.~\eqref{eaS} as the dominant contribution in the saddle-point approximation of path integrals.

The simplest cosmological spacetime is given by the spatially-flat Friedmann-Lemaitre-Robertson-Walker (FLRW) metric:
\beq
\d s^2
=
-N^2 \,\d t^2
+ a^2 \left[(\d x^1)^2 + (\d x^2)^2 + (\d x^3)^2\right]
\,
\label{FLRW}
\eeq
where $N=N(t)$ denotes the lapse function and $a=a(t)$ is the scale factor. Without loss of generality, we have set the shift functions $N_i$ to zero. For a homogeneous scalar field $\phi = \phi(t)$, we find
\begin{align}
a^3_\pm(t)
&=
\pm 3 \sqrt{\kappa} \, p_\phi \, t
\label{asol}
\\
\phi_\pm(t)
&=
\pm \frac{1}{\sqrt{\kappa}}\,
\log\left(\pm \frac{t}{t_0}\right)
\ ,
\label{psol}
\end{align}
where $t_0$ is an integration constant, $p_\phi = a^3\, \dot{\phi}$ is a constant of motion
that follows from the equation for $\phi$ and we have set $N=1$ in the final expressions.
The different signs above correspond to different regimes of evolution of the universe. Expansion takes place for the positive sign, with $0 < t < \infty$, and contraction for the negative sign, with $-\infty < t < 0$.
We have also adjusted the integration constants accordingly in order to obtain $a_\pm(0) = 0$.
With such a choice we can join the two regimes of evolution at $t=0$ to form a ``bouncing''
configuration, which shall be denoted by $\varphi^i_{\rm s}=(a_{\rm s}(t),\phi_{\rm s}(t))$.

It is not difficult to show that the Ricci scalar for the solution~\eqref{asol} diverges for $t\to 0$,
which indicates the existence of a spacetime singularity at the bounce. We also note that the determinant of the spacetime metric vanishes at the bounce, which could suggest the presence of a covariant singularity for $\epsilon > 0$ (see Eq.~\eqref{eq:general}). The effective action~\eqref{eaS} diverges when evaluated on the solutions~\eqref{asol}-\eqref{psol} for $t\to 0$. Thus, everything seems to indicate that the spacetime singularity at $t=0$ corresponds to a functional singularity, which would prevent us from defining observables for the bouncing solution $\varphi^i_{\rm s}$.

However, the calculation of the functional winding number $\mathcal W$ shows otherwise. Following the formalism of Sec.~\ref{Ssingularity}, we encircle the potentially singular configuration $\varphi_s^i$ with a curve $\gamma_1$ parameterised as
\beq
\gamma^I(t;\theta)
=
\left(a_{\rm s}(t) + A\,\cos\theta, \phi_{\rm s}(t) + A \,\sin\theta, 1\right)
\ ,
\label{parS}
\eeq
for all values of $t$ for which $a=a_{\rm s}(t)$ and $\phi=\phi_{\rm s}(t)$ are defined, and $A$ is a positive constant.
Since the effective action diverges when computed along \eqref{parS}, the calculation of the winding number for such a parameterization is
quite tricky. Nothing forbids us from parameterizing $\gamma_1$ differently, but it is easier to exploit the freedom to add total derivatives
to the effective action. By including the total derivative
\beq
F
=
\frac{A^3}{3}\, \cos^3\theta\, \dot \phi + \frac{A^2}{2}\, \cos^2\theta \,\dot a
\ ,
\eeq
we can cancel out the divergence in the time integral over the
configurations~\eqref{parS}. This results in $\Gamma[\varphi_{\rm s}]=0$, already suggesting that the apparent singularity is removable. 
Indeed, the effective action evaluated along~\eqref{parS} vanishes identically, namely $\Gamma(\theta) = 0$, which yields
\begin{equation}
	\mathcal W = 0
	\ .
\end{equation}
This implies that the apparent singularity at $\varphi_{\rm s}$ is indeed removable by local alterations of the effective action in the vicinity of $\varphi_{\rm s}$. In fact, by imposing a cutoff $T>0$ in the lower limit of the time integral in Eq.~\eqref{eaS} and taking $T\to 0$ in the end, one finds
\begin{equation}
	\lim\limits_{T\to 0}
	\Gamma_T[\varphi_{\rm s}]=0
	\ ,
\end{equation}
where $\Gamma_T[\varphi_{\rm s}]$ denotes the regularized effective action.
Therefore, the spacetime singularity at $t=0$ does not correspond to a functional singularity and physical observables can be defined normally.
This shows that configuration-space coordinates must exist in which the spacetime singularity vanishes
completely~\cite{Kamenshchik:2016gcy, Kamenshchik:2017ojc, Kamenshchik:2018crp}.
\section{Conclusions}
\label{Sconc}
\setcounter{equation}{0}
In this article, we have reviewed some aspects of covariant singularities. It has been known for some time now that some spacetime singularities can be removed by field redefinitions. This is reflected in the fact that the Hawking-Penrose theorem does not transform covariantly under such transformations. The authors have thus proposed the idea of covariant singularities, namely singularities that are invariant under both spacetime and configuration-space diffeomorphisms. Such singularities can be envisaged as boundary points in the infinite-dimensional configuration space or configurations at which the Vilkovisky-DeWitt effective action is undefined, the so-called functional singularities. We argued that the former can be interpreted as missing configurations, which offers no issues to physics as they cannot be solutions to the theory. The latter, on the other hand, are solutions by definition, but no observables may be defined for them. Functional singularities are of topological origin and affect the entire structure of the theory, thus they cannot be removed without severely modifying the physical model.

We have shown a simple example in cosmology where a spacetime singularity is known to exist, but no corresponding functional singularities take place. Despite the apparent spacetime singularity, observables can be well-defined in this case. There are, indeed, different configuration-space coordinates that make the aforementioned spacetime singularity look smooth. There is, however, no guarantee that functional singularities are absent for all popular physical models. While there remains a lot to be learned about these covariant singularities, the functional winding number provides a systematical procedure for building self-consistent models, which is the first step for the construction of a quantum theory of gravity.
\subsection*{Acknowledgments}
I.K., A.K. ~and R.C.~are partially supported by the INFN grant FLAG.
The work of R.C.~has also been carried out in the framework of activities
of the National Group of Mathematical Physics (GNFM, INdAM).
A.K.~is partially supported by the Russian Foundation for Basic Research 
grant No  20-02-00411.
I.K.~was also supported by the Conselho Nacional de Desenvolvimento Cient\'ifico e Tecnol\'ogico (grant no.~162151/2020-9).

\end{document}